\documentclass[aps,pra,superscriptaddress,amsmath,amssymb,preprintnumbers
,twocolumn,showpacs]{revtex4}

\usepackage{graphicx}

\usepackage{dcolumn}

\usepackage{bm}

\usepackage{amsmath}

\begin{document}


\title{\LARGE \bf Stationary entanglement induced by dissipation }

\author{S. Nicolosi}
\author{A. Napoli}
\author{A. Messina}
\address{INFM, MIUR and Dipartimento di Scienze Fisiche ed
Astronomiche, Universit\`{a} di Palermo, via Archirafi 36, 90123
Palermo, Italy, Tel/Fax: +39 91 6234243; E-mail:
messina@fisica.unipa.it }
\author{F. Petruccione}
\address{School of Pure and Applied Physics University of KwaZulu-Natal Durban, 4041,
South Africa and Istituto
  Italiano per gli Studi Filosofici, Via Monte di Dio 14, I-80132
  Napoli, Italy}
\begin{abstract}
The dynamics of two two-level dipole-dipole interacting atoms
coupled to a common electromagnetic bath and closely located
inside a lossy cavity, is reported.  Initially injecting only one
excitation in the two atoms-cavity system, loss mechanisms
asymptotically drive the matter sample toward a stationary
maximally entangled state.  The role played by the closeness of
the two atoms with respect to such a cooperative behaviour is
carefully discussed. Stationary radiation trapping effects are
found and transparently interpreted.
\end{abstract}

\pacs{03.65.Yz, 03.67.Mn, 42.50.Fx}

\maketitle

\medskip
\pagebreak A system prepared in an entangled state may show rather
puzzling behavior and counterintuitive features. In such a state,
for example, the system exhibits the astonishing property that the
results of measurements on one subsystem cannot be specified
independently from the parameters of the measurements on the other
components.
The renewed and more and more growing interest toward entanglement
concept reflects the consolidated belief that unfactorizable
states of multipartite system provide an unreplaceable applicative
resource, for example, in the quantum computing  research area
\cite{Nielsen}. However, the realization of quantum computation
protocols suffers from the difficulty of isolating a quantum
mechanical system from its environment. Unavoidably, the
system-environment interaction leads to decoherence phenomena
which, as intuition suggests, are always noxious for quantum
computers, since they imply loss of the information stored in the
system. This circumstance is at the origin of an intense research
aimed at proposing theoretical and experimental schemes to fight
or control decoherence manifestations
\cite{Cinesi,Cinesi2,Cirac,Shneider,Yang,Hagley,Foldi}. Very
recently, however, the possibility of environment-induced
entanglement in an open quantum system has opened new intriguing
perspectives \cite{Ficek02,Beige99,Pellizzari,Benatti}. For
example, it has been shown that dissipation can be exploited to
implement nearly decoherence-free quantum gates
\cite{Beige03,BeigePRA03}, the main requirement being the
existence of a decoherence-free subspace for the system under
consideration \cite{Palma}. Moreover, despite the widely held
belief, it has been shown that transient entanglement between
distant atoms can be induced by atomic spontaneous decay or by
cavity losses \cite{Beige99,Benatti,Ficek03}. It has been also
demonstrated \cite{Jakobczyk} that asymptotic (stationary)
entangled states of two closely separated two-level atoms in  free
space can be created by spontaneous emission. Notwithstanding the
nearness of the two atoms, dipole-dipole interaction is however
neglected in Ref. \cite{Jakobczyk}. In this paper we investigate
the dynamics of a couple of spontaneously emitting two-level atoms
confined within a bad single-mode cavity, taking into account from
the very beginning their dipole-dipole interaction. Our main
result is that in such a condition the matter subsystem, even
experiencing a further relatively faster energy loss mechanism,
may be as well conditionally guided toward a stationary robust
entangled state. We shall prove that this state  is the
antisymmetric one with respect to the exchange of the two
two-level atoms provided that all the initial energy given to the
system is concentrated in the matter subsystem only.

Let us suppose that two identical two-level atoms are located at
$\mathbf{r}_1$ and $\mathbf{r}_2$ inside a resonant single mode
cavity. Putting $R=\vert
\mathbf{r}_1-\mathbf{r}_2\vert\equiv\vert\mathbf{R}\vert$ and
denoting by $\theta$ the angle between $\mathbf{R}$ and the atomic
transition dipole moment $\mathbf{d}$, the Hamiltonian describing
the dipole interaction between the two atoms can be written in the
form \cite{Leonardi,Leonardi90}
\begin{equation}\label{dip-dip}
   H_{12}=\hbar \eta (\sigma_+^{(1)}\sigma_-^{(2)}+h.c.),
\end{equation}
where $\eta= \frac{3}{4}\frac{\Gamma_0c^3}{\omega_0^3 R^3}
  (1-3cos^2\theta)$ and $\Gamma_0$ is the spontaneous
emission rate in free space. In eq.\ (\ref{dip-dip})
$\sigma_{\pm}^{(i)}$ $(i=1,2)$ are the Pauli operators of the
$i$-th atom. Assume in addition that all the conditions are
satisfied under which the interaction between each atom and the
cavity field is described by a Jaynes Cummings (JC) model
\cite{JC}. Under these hypotheses, the unitary time evolution of
the system under scrutiny is governed by the following Hamiltonian
\begin{equation}\label{HAC}
  H_{AC}=\hbar \omega
  \alpha^{\dag}\alpha+\frac{\hbar\omega_0}{2}\sum_{i=1}^2\sigma_z^{(i)}+
  \hbar\sum_{i=1}^2[\varepsilon^{(i)} \alpha\sigma_+^{(i)}+h.c.]+H_{12},
\end{equation}
where $\hbar\omega_0$ denotes the energy separation between the
$i$-th atom $(i=1,2)$ excited $(\vert + \rangle_i)$ and ground
$(\vert -\rangle_i )$ states, $\omega\sim\omega_0$ is the
frequency of the cavity mode and $\varepsilon^{(i)}$ is the $i-$th
atom-field mode coupling constant. In the above equation $\alpha$
($\alpha^{\dag}$) is the annihilation (creation) operator relative
to the single-mode cavity field and $\sigma_z^{(i)}$ the $i-$th
atom inversion operator. It is easy to demonstrate that the total
excitation number operator
$\hat{N}=\alpha^{\dag}\alpha+\frac{1}{2}(\sigma_z^{(1)}+\sigma_z^{(2)})+1$
is a constant of motion. Thus, preparing the physical system at
$t=0$ in a state with a well defined number of excitations $N_e$,
its dynamics is confined in the finite-dimensional Hilbert
subspace singled out by this eigenvalue of $\hat{N}$. In a
realistic situation, however, the system we are considering
evolves under the action of different sources of decoherence.
First, photons can leak out through the cavity mirrors due to the
coupling of the resonator mode to the free radiation field.
Moreover, the atoms confined in the resonator can spontaneously
emit photons into non-cavity field modes. The microscopic
Hamiltonian which takes into account all these loss mechanisms can
be written in the form \cite{Haroche}
\begin{equation}\label{Htot}
  H=H_{AC}+H_R+H_{AR}+H_{CR},
\end{equation}
where $H_{AC}$ is given by Eq.\ (\ref{HAC}),
\begin{equation}\label{HR}
 H_R=\hbar\sum_{\mathbf{k},\lambda}\omega_{\mathbf{k}, \lambda}[c_{\mathbf{k}, \lambda}^{\dag}c_{\mathbf{k}, \lambda}
 + \tilde{c}_{\mathbf{k}, \lambda}^{\dag} \tilde{c}_{\mathbf{k}, \lambda}]
\end{equation}
is the Hamiltonian of the environment,
\begin{equation}\label{HAR}
  H_{AR}=\sum_{\mathbf{k},\lambda,i}[g_{\mathbf{k},\lambda}^{(i)}\tilde{c}_{\mathbf{k},\lambda}\sigma_+^{(i)}+h.c.]
\end{equation}
describes the interaction between the atoms and the bath and,
finally,
\begin{equation}\label{HCR}
  H_{CR}=\sum_{\mathbf{k},\lambda}[s_{\mathbf{k},\lambda}c_{\mathbf{k},\lambda}\alpha^{\dag}+h.c.]
\end{equation}
takes into account the coupling between the environment and the
cavity field. As usual, in Eq.\ (\ref{HR}) we have assumed that
the two subsystems,  the two atoms and the single-mode cavity, see
two different reservoirs in the following both assumed at $T=0$.
The boson operators relative to the atomic bath are denoted by $\{
\tilde{c}_{\mathbf{k},\lambda},
\tilde{c}_{\mathbf{k},\lambda}^{\dag}\}$ whereas
$c_{\mathbf{k},\lambda}, c_{\mathbf{k},\lambda}^{\dag}$ are the
$(\mathbf{k},\lambda)$ mode annihilation and creation operators
respectively of the cavity environment. Moreover, the coupling
constants $s_{\mathbf{k},\lambda}$ are phenomenological parameters
and
\begin{equation}\label{gklambda}
  g_{\mathbf{k},\lambda}^{(i)}=-i(\frac{2\pi\hbar\omega_0^2}{V\omega_k})
  ^\frac{1}{2} \mathbf{e}_{k\lambda} \cdot \mathbf{d}e^{i \mathbf{k} \cdot
  \mathbf{r}_{i}}
\end{equation}
stems from a dipole atom field coupling \cite{Leonardi}. In Eq.\
(\ref{gklambda}) $\mathbf{e}_{\mathbf{k}\lambda}$ represents the
polarization vector of the thermal bath $(\mathbf{k} \lambda)$
mode of frequency $\omega_\mathbf{k}$ and $V$  its effective
volume. In the Rotating-Wave and Born-Markov approximations
\cite{F.Petruccione,W.H.Louiselle}, the reduced density operator
$\rho_{AC}$  relative to the bipartite system composed by the two
atoms subsystem and the single-mode cavity, evolves  nonunitarily
in accordance with the following quantum master equation in
Lindblad form:
\begin{equation}\label{m-e}
  \dot{\rho}_{AC}=-\frac{i}{\hbar} [H_{AC},\rho_{AC}]+
  \mathcal{L}_f\rho_{AC}+ \mathcal{L}_A\rho_{AC},
\end{equation}
where
\begin{equation}\label{Lf}
 \mathcal{L}_f\rho_{AC}=
k(2\alpha\rho_{AC}\alpha^{\dag}-\alpha^{\dag}\rho_{AC}\alpha-\rho_{AC}\alpha^{\dag}\alpha),
\end{equation}
and
\begin{eqnarray}\label{La}
&&\mathcal{L}_A\rho_{AC} = \\ \nonumber
 &&\sum_{i=1}^{2}{\Gamma(2\sigma_{-}^{(i)}\rho_{AC}\sigma_{+}^{(i)}-
 \sigma_{+}^{(i)}\sigma_{-}^{(i)}\rho_{AC}-\rho_{AC}\sigma_{+}^{(i)}\sigma_{-}^{(i)})}
 + \\ \nonumber &&\Gamma_{12}(2\sigma_{-}^{(1)}\rho_{AC}\sigma_{+}^{(2)}-
  \sigma_{+}^{(1)}\sigma_{-}^{(2)}\rho_{AC}-\rho_{AC}\sigma_{+}^{(1)}\sigma_{-}^{(2)})
 + \\ \nonumber &&\Gamma_{21}(2\sigma_{-}^{(2)}\rho_{AC}\sigma_{+}^{(1)}-
 \sigma_{+}^{(2)}\sigma_{-}^{(1)}\rho_{AC}-\rho_{AC}\sigma_{+}^{(2)}\sigma_{-}^{(1)}).
\end{eqnarray}
In the above equations we have introduced
$\Gamma=\frac{4\pi\omega_0^3|\mathbf{d}|^2}{3 \hbar c^3}$ and
$\Gamma_{12}\equiv\Gamma_{21}=\Gamma f(R)$ where \cite{Agarwal}
\begin{eqnarray}\label{f}
&&f(R)=\frac{3}{2}\{[1-(cos \vartheta)^2]c\frac{\sin(\frac{\omega_0}{c}R)}{\omega_0R}+ \\
\nonumber &&[1-3(cos
\vartheta)^2][c^2\frac{\cos(\frac{\omega_0}{c}R)}{(\omega_0R)^2}-c^3\frac{\sin(\frac{\omega_0}{c}R)}
{(\omega_0R)^3}] \}.
\end{eqnarray}
with $0\leq f(R)\leq 1$. We emphasize that $\Gamma_{12}$ measures
the strength of the atom-atom cooperation induced by their
coupling with a common bath. Finally,
$k=\sum_{\mathbf{k}\lambda}{\vert s_{\mathbf{k}\lambda}\vert ^2
  \delta(\omega_{\mathbf{k}}-\omega)}$, appearing
in Eq.\ (\ref{Lf}), is the cavity decay rate coefficient. Assume
now that the two atoms inside the cavity are closely located that
is $c<<R\omega_0$. In such a situation the cooperation between the
two atoms stemming from their interaction with a common bath, is
maximum that is $f(R)=1$. Moreover it is reasonable to put
$\varepsilon^{(1)}=\varepsilon^{(2)}\equiv \varepsilon$. We are
able to exactly solve Eq.\ (\ref{m-e}) when this point-like model
is adopted.

\begin{figure}
 \hspace{0.5 cm}
 \includegraphics[width=8 cm,height=10 cm]{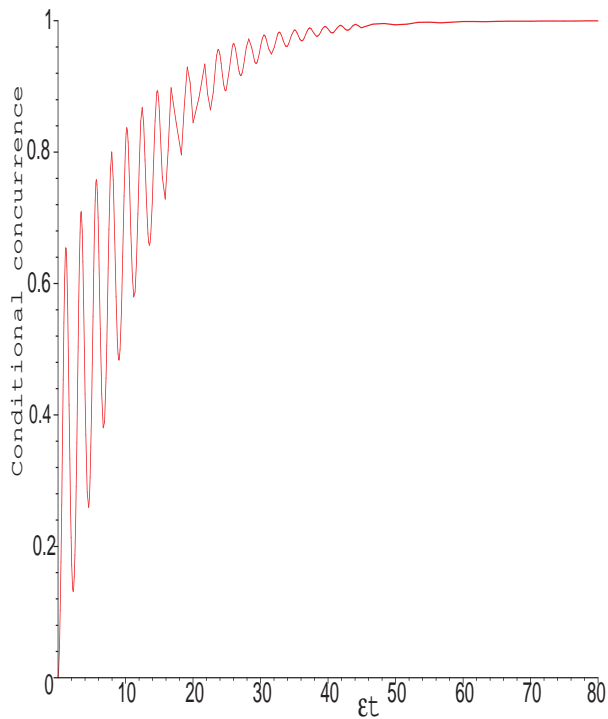}
  \caption{ \footnotesize Conditional Concurrence $C$ in correspondence to $\eta=0.5\varepsilon$,
  $k=10^{-1}\varepsilon$ and $\Gamma=10^{-2}\varepsilon$}
  \label{Conc}
\end{figure}

To this end consider the unitary operator
 $U$ defined as
\begin{equation}\label{U}
  U=\exp [{-\frac{\pi}{4}(\sigma_+^1\sigma_-^2-\sigma_-^1\sigma_+^2)}].
\end{equation}
It can be shown that $[U,\hat{N}]=0$ and that, confining our
attention on the Hilbert subspace correspondent to $N_e=0,1$, Eq.\
(\ref{m-e}) can be equivalently cast in the form
\begin{eqnarray}\label{m-e-tr}
  \dot{\tilde{\rho}}_{AC}&\equiv&U^{\dag}\rho_{AC} U=-\frac{i}{\hbar}
  [\tilde{H}_{AC},\tilde{\rho}_{AC}]\\ \nonumber
  &+&k(2 \alpha\tilde{\rho}_{AC}\alpha^{\dag}-
  \alpha^{\dag}\alpha\tilde{\rho}_{AC}-\tilde{\rho}_{AC}\alpha^{\dag}\alpha)\\
  \nonumber
  &+&2 \Gamma(2 \sigma_-^{(1)}\tilde{\rho}_{AC}\sigma_+^{(1)}-
  \sigma_+^{(1)}\sigma_-^{(1)}\tilde{\rho}_{AC}-\tilde{\rho}_{AC}\sigma_+^{(1)}\sigma_-^{(1)}),
\end{eqnarray}
where $\tilde{\rho}_{AC} \equiv U^{\dag}\tilde{\rho}_{AC}U$
\begin{eqnarray}\label{HACtras}
 \tilde{H}_{AC}&&\equiv  U^{\dag}H_{AC}U=  \hbar \omega
  \alpha^{\dag}\alpha+\hbar\omega_0\sum_{i=1}^2\sigma_z^{(i)}+\\ \nonumber
 && \hbar \varepsilon_{eff}[\alpha\sigma_+^{(1)}+h.c.]-
  \frac{\hbar \eta}{2}
  (\sigma_z^{(2)}-\sigma_z^{(1)}),
\end{eqnarray}
with $\varepsilon_{eff}=\sqrt{2}\varepsilon$. It is important to
notice that in the new representation the dipole-dipole
interaction  $H_{12}$  given by Eq.\ (\ref{dip-dip}), only
renormalizes the atomic frequencies. The transformed Hamiltonian
$\tilde{H}_{AC}$  puts transparently into evidence that the system
of two atoms cooperates in the interaction with the cavity field.
In this new representation two two-level fictitious atoms appear,
only one of them being coupled, by means of a simple JC
interaction model, to the cavity field. The circumstance that the
atomic sample can exchange energy with the field through only one
of its collective atoms, provides a natural explanation for the
radiation trapping phenomena \cite{Benivegna88}. This conclusion
is true under ideal conditions when neither cavity losses or
atomic spontaneous decay are considered. Thus, it seems
interesting to investigate whether such an energy storage
mechanism survives in a more realistic situation like the one
under scrutiny in this paper. Looking at Eq.\ (\ref{HACtras}) we
may see also that in the transformed representation the atomic
subsystem looses its energy only through the interaction of the
first atom with both the cavity mode and the environment. Such a
behaviour stems from the fact that the other atom freely evolves
being decoupled from either the cavity field and  the
electromagnetic modes of the thermal bath. Comparing this result
with the interpretation of the energy trapping given in Refs.
\cite{Benivegna88,Benivegna89}, we may immediately catch the main
role played by the closeness of the two atoms in our model. It is
indeed just this feature which, in the transformed representation,
leads to the existence of one atom immune from spontaneous
emission losses and, at the same time, decoupled from the cavity
mode. Thus, to locate the atomic sample within a linear dimension
much shorter than the wavelength of the cavity mode, introduces a
permutational atomic symmetry which is at the origin of a
collective behaviour of the two atoms. {\sl As a direct
consequence the matter system may stationarily trap the initial
energy even in presence of both the proposed dissipation
channels}. Suppose that only one excitation is initially present
in the atomic subsystem whereas the cavity is prepared in its
vacuum state. From an experimental point of view it seems
reasonable to think that the excitation given to the matter system
is captured by the atom 1 or by the atom 2 with the same
probability. On the other hand, exciting only an a priori fixed
atom or preparing a quantum coherent superposition of the two
states $\vert + \rangle_1 \vert - \rangle_2$ and $\vert -
\rangle_1 \vert + \rangle_2$, could present serious practical
difficulties. Stated another way, our initial condition can be
reasonably expressed by $\rho_{AC}(0)=\rho_A(0)\otimes \rho_C(0)$
with $\rho_A=\frac{1}{2}(\vert + \rangle_1 \vert - \rangle_{2 \;
1} \langle + \vert_2 \langle - \vert +  \vert - \rangle_1 \vert +
\rangle_{2 \; 1} \langle - \vert_2 \langle + \vert)$ and
$\rho_C(0)=\vert 0 \rangle \langle 0 \vert$, $\vert 0 \rangle$
being  the vacuum Fock state of the cavity mode. It is evident
that, under this hypothesis, at a generic time instant $t$, the
density operator $\rho_{AC}$, can have non zero matrix elements
only in the Hilbert subspace generated by the following ordered
set of four state vectors
\begin{equation}\label{basis}
  \{ \vert 0 \rangle \vert - \rangle_1 \vert - \rangle_2 ;
  \vert 0 \rangle \vert + \rangle_1\vert - \rangle_2 ;\vert 0 \rangle \vert -
  \rangle_1\vert + \rangle_2;
  \vert 1 \rangle\vert - \rangle_1 \vert - \rangle_2\}.
\end{equation}
Taking into account that $[U,\hat{N}]=0$, the same conclusion
holds
 for $\tilde{\rho}_{AC}$ too. This fact provides the key to solve
exactly eq.\ (\ref{m-e-tr}). One finds
\begin{equation}\label{matrix1}
\tilde{\rho}_{AC}(t)=
\begin{pmatrix}
  \tilde{\rho}_{11}(t) & 0 & 0 & 0 \\
  0 & \tilde{\rho}_{2,2}(t) & 0 & \tilde{\rho}_{2,4}(t) \\
  0 & 0 & \tilde{\rho}_{3,3}(t) & 0 \\
  0 & \tilde{\rho}_{2,4}^*(t) & 0 & \tilde{\rho}_{4,4}(t)
\end{pmatrix},
\end{equation}
where $\tilde{\rho}_{3,3}(t)=\tilde{\rho}_{3,3}(0)$,
$\tilde{\rho}_{11}(t)=1-\sum_{i=2}^{4}\tilde{\rho}_{ii}(t)$ with
\begin{eqnarray}\label{rho22}
\nonumber && \tilde{\rho}_{22}(t)=
\frac{e^{-A_+t}}{\Omega_1^2+\Omega_2^2}\{(\vert\Delta\vert^2+\frac{\Omega_2^2+\Omega_1^2}{4})
\cosh(\Omega_1 t)\\
\nonumber &&
+(\frac{\Omega_2^2+\Omega_1^2}{4}-\vert\Delta\vert^2)\cos(\Omega_2
t)+(\frac{\Omega_2A_-}{2}-\frac{\Omega_1\eta}{2})\\
&&\sin(\Omega_2
t)+(\frac{\Omega_2}{2}\eta+\frac{\Omega_1}{2}A_-)\sinh(\Omega_1
t))\}
\end{eqnarray}
\begin{equation}\label{rho44}
\tilde{\rho}_{44}(t)=\frac{\varepsilon^2_{eff}}{\Omega_1^2+\Omega_2^2}(\cosh(\Omega_1
t)-\cos(\Omega_2 t))e^{-A_+t}
\end{equation}
\begin{eqnarray}\label{rho24}
\nonumber &&\tilde{\rho}_{2,4}(t)=
\frac{e^{-A_+t}}{\Omega_1^2+\Omega_2^2}\{
\frac{i\varepsilon_{eff}}{2}(\Omega_2-i\Omega_1)(\sin(\Omega_2 t)+\\
&& -i \sinh(\Omega_1 t))+\Delta \varepsilon_{eff}(\cosh(\Omega_1
t)-\cos(\Omega_2 t))\}
\end{eqnarray}
where $A_{\pm}=k \pm 2\Gamma$, $2\Delta=\eta+iA_-$
\begin{equation}\label{xiOmega}
\Omega_i=[(-1)^i\frac{P}{2}+\frac{1}{2}(P^2-4V)^{1/2}]^{1/2}
\end{equation}
with $P=\eta^2+4\varepsilon^2_{eff}-A_-^2$ and $V=-A_-^2\eta^2$
and $\omega=\omega_0$. We emphasize that, on the basis of the
block diagonal form exhibited by $\tilde{\rho}_{AC}(t)$,  the
transformed system, at a generic time instant $t$, is in a
statistical mixture of the vacuum state $\vert 0 \rangle \vert -
\rangle_1 \vert - \rangle_2$ and of a one excitation appropriate
density matrix describing with certainty the storage of the
initial energy.  Taking into account the easily demonstrable
inequality $\Omega_1<A_+$, it is immediate to convince oneself
that for { \sl  $t\gg A_+^{-1}$ the correspondent asymptotic form
assumed by $\tilde{\rho}_{AC}$ is time independent and such that
the probability of finding energy in the effective JC subsystem
exactly vanishes}. Considering that the initial condition imposed
on $\rho_{AC}(0)$ may be converted into
$\tilde{\rho}_{ii}=\frac{1}{2}(\delta_{i2}+\delta_{i3})$ and
$\tilde{\rho}_{ij}=0$ for $i\neq j$, we may conclude that at $t\gg
A_+^{-1}$ the only two matrix elements different from zero are
$\tilde{\rho}_{11}(t)\equiv\tilde{\rho}_{22}(0)$ and
$\tilde{\rho}_{33}(t)\equiv\tilde{\rho}_{33}(0)$. Transforming
$\tilde{\rho}_{AC}(t)$ back to the original representation, the
exact solution $\rho_{AC}$ for the reduced density matrix of the
system under scrutiny is then easily found. Since the unitary
operator $U$ is time independent, we are legitimated to forecast
an asymptotic time independent solution  in the original
representation too. The reduced density matrix can indeed be
written in the compact form
\begin{equation}\label{roAC}
  \rho_{AC}=\frac{1}{2}\vert \psi_D\rangle\langle \psi_D
  \vert+\frac{1}{2}\vert \psi_T\rangle\langle \psi_T
  \vert,
\end{equation}
with $ \vert \psi_D \rangle=\vert 0 \rangle \vert - \rangle_1
\vert -
  \rangle_2$ and
\begin{equation}\label{psiT}
  \vert \psi_T \rangle = \frac{1}{2}[ \vert - \rangle_1 \vert +
  \rangle_2 - \vert + \rangle_1\vert -
  \rangle_2].
\end{equation}
eigenstate of $( \vec{\sigma}_1+\vec{\sigma}_2)^2$ with eigenvalue
zero. Eq.\ (\ref{roAC}) suggests that $stationary$ entangled
states of the two atoms can be generated putting outside of the
cavity single photon detectors allowing a continuous monitoring of
the decay of the system through the two possible channels (atomic
and cavity dissipation). Reading out the detectors states at
$\bar{t} \gg A_+^{-1}\approx k^{-1}$, if no photon has been
emitted, then as a consequence of this measurement outcome, our
system is projected into the state $\vert \psi_T \rangle$ given by
Eq.\ (\ref{psiT}). {\sl This is the main result of our paper which
means that a successful measurement, performed at large enough
time instants, generates an uncorrelated state of the two
subsystems atoms and cavity, leaving the atomic sample in its
maximally antisymmetric entangled state}\ (\ref{psiT}). To analyze
the time evolution of the degree of entanglement that gets
established between the two initially uncorrelated atoms, we
exploit the concept of concurrence $C$ first introduced by
Wootters \cite{Wootters97, Wootters98}. If at an assigned time $t$
no photons have been emitted, the {\sl conditional concurrence}
$C$ assumes the form
\begin{equation}
 C(t)=\frac{\vert\tilde{\rho}_{22}(t)-\tilde{\rho}_{33}(t)\vert}
 {\tilde{\rho}_{22}(t)+\tilde{\rho}_{33(t)}+\tilde{\rho}_{44}(t)}
\end{equation}
As clearly shown in Fig. \ref{Conc} the degree of entanglement
between the two atoms, starting from zero, reaches its maximum
value, that is $C=1$, in a time of the order of $k^{-1}$, after a
large number of oscillations.

Summarizing, in this paper we have analyzed the dynamics of a
realistic system composed of two two-level atoms with
dipole-dipole interaction, embedded in a bad single mode cavity
and coupled to a common electromagnetic environment. The exact
analytic solution of the Markovian dissipative dynamics reveals
that the environment induces stationary entangled states of the
two atoms starting from a realistic preparation of the matter
subsystem. This fact makes it possible the experimental
realization of this process with real atoms and could be of some
relevance for the development of quantum devices.

\end{document}